# Human or AI? Comparing Design Thinking Assessments by Teaching Assistants and Bots

Sumbul Khan
Science, Mathematics and Technology
Singapore University of Technology and Design
Singapore
sumbul_khan@sutd.edu.sg
ORCID: 0000-0003-4617-0155

Wei Ting Liow
Science, Mathematics and Technology
Singapore University of Technology and Design
Singapore
weiting_liow@sutd.edu.sg
ORCID: 0000-0001-7122-1722

Lay Kee Ang
Science, Mathematics and Technology
Singapore University of Technology and Design
Singapore
ricky_ang@sutd.edu.sg
ORCID: 0000-0003-2811-1194

***Abstract*—As design thinking education is growing in secondary and tertiary education, educators face a mounting challenge of evaluating creative artefacts that comprise visual and textual elements. Traditional, rubric-based methods of assessment are laborious, time-consuming, and inconsistent, due to their reliance on Teaching Assistants (TAs) in large, multi-section cohorts. This paper presents an exploratory study to investigate the reliability and perceived accuracy of AI-assisted assessment vis-à-vis TA-assisted assessment in evaluating student posters in design thinking education. Two activities were conducted with 33 Ministry of Education (MOE), Singapore school teachers, with the objective (1) to compare AI-generated scores with TA grading across three key dimensions: empathy and user understanding, identification of pain points and opportunities, and visual communication, and (2) to understand teacher preferences for AI-assigned, TA-assigned, and hybrid scores. Results showed low statistical agreement between instructor and AI scores for empathy and pain points, though slightly higher alignment for visual communication. Teachers generally preferred TA-assigned scores in six of ten samples. Qualitative feedback highlighted AI's potential for formative feedback, consistency, and student self-reflection, but raised concerns about its limitations in capturing contextual nuance and creative insight. The study underscores the need for hybrid assessment models that integrate computational efficiency with human insights. This research contributes to the evolving conversation around responsible AI adoption in creative disciplines, emphasizing the balance between automation and human judgment for scalable and pedagogically sound assessment practices.**

## I. INTRODUCTION

Design thinking is a human-centered approach to innovation that draws from the designer's toolkit to integrate the needs of people, the possibilities of technology, and the requirements for business success. It is a non-linear, iterative process that teams use to understand users, challenge assumptions, redefine problems, and create innovative solutions to prototype and test. Design thinking is particularly useful for tackling ill-defined or unknown problems and involves five phases: Empathize, Define, Ideate, Prototype, and Test.

Design thinking education is increasingly being integrated into secondary and tertiary curricula, often through project-based learning and interdisciplinary modules. In Singapore, one of the mandatory subjects for 13- and 14-year-old students at lower secondary is Design and Technology (D&T), which emphasizes design thinking through user research, sketching, ideation, and the development of design solutions [1]. Each year, approximately 37,208 Secondary 1 students are enrolled in Singapore's secondary schools, and all are required to take D&T, which incorporates key design thinking practices [2]. A typical cohort of students enrolled in the Design Thinking module at a secondary school in Singapore usually ranges from 250 to 400 students.

In design thinking programs, students often present their work as posters that blend visual elements with textual content, showcasing components of the design thinking process such as empathy maps, personas, user research analyses, affinity diagrams, clusters of ideas, and concept directions.

Assessing visual content, such as posters, presents unique challenges. Unlike written responses, posters rely heavily on layout, structure, and visual storytelling to convey process thinking, insights, and proposed solutions. As design thinking programs scale, instructors face mounting challenges in assessing student work in design programs. While there has been some research in the assessment of abstract qualities such as empathy, critical thinking, and creativity, the evaluation of these qualities remains difficult and time-consuming for educators [3].

Manual assessment methods using rubrics, while pedagogically rich, are time-consuming, labor-intensive, and prone to inconsistency, especially when multiple assessors are involved in large cohorts with multiple sections [4]. This complexity is further compounded when multiple raters conduct assessments, each potentially interpreting rubrics differently. In large-scale courses, the assessment of student artefacts often relies heavily on teaching assistants (TAs). With class sizes frequently exceeding 300 students, instructors delegate grading responsibilities to TAs or Assistant teachers to manage workload and ensure timely feedback.

Differences in perception and rubric interpretation between assessors can lead to inconsistencies. Studies have argued that it is challenging to accommodate the subjective and creative nature of visual design, making the justification of assessment decisions difficult [5]. In previous studies, we found that even expert assessors diverge in scoring when textual explanations are absent, and novice assessors struggle with consistency despite rubric support [6].

Researchers have identified five critical limitations in traditional educational assessment systems: (1) they are burdensome for educators, (2) offer limited insight into student learning, (3) fail to adapt to student differences, (4) emphasize school-based norms that may not connect to authentic, real-world applications, (5) they often focus on outdated skills, many of which can now be performed by machines, reducing their relevance in today's digital environment [7]. These criticisms apply directly to design thinking, which is dynamic, user-centred, and context-specific.

The context of this study is a multi-section classroom environment where multiple assessors are involved. Teachers in these large classes are heavily reliant on Teaching Assistants to help manage the substantial grading workload. In such settings, a coordinator can deploy an AI-based grading system, such as a custom GPT, using a provided rubric and example responses. This exploratory study aimed to understand the reliability and perceived accuracy of AI-assisted versus TA-assisted assessment in evaluating student works, such as posters that comprise both textual and visual elements. The main contributions of this paper are

1. We present findings of comparison of AI-assigned scores and TA-assigned scores with Instructor-assigned benchmark scores for three different dimensions, namely (a) Empathy and user understanding, (b) Identification of pain points and opportunities, and (c) Visual communication.
2. We present findings of teachers' perceptions of AI-assigned, TA-assigned, and hybrid scores.
3. We present qualitative findings of secondary school teachers' stance on AI-based assessment of such students' works, such as posters, that comprise both textual and visual content.
4. Based on the above, we provide insights on teachers' readiness to adopt AI graders for such assessment tasks.

## II. LITERATURE

Among the current approaches for evaluating creative work, using rubrics is still the most widely adopted and practical way to assess design thinking assignments. Rubrics provide a structured framework for assessing dimensions such as empathy, ideation, originality, and clarity. A recent study advocates for performance-based rubrics that capture multi-dimensional aspects of critical and creative thinking [8]. A previous study developed and validated Scale of Design Thinking for Teaching (SDTT), a rubric-based framework that focuses on phases such as problematizing, ideating, prototyping, and testing [9]. Although designed for teachers, its conceptual model aligns closely with design thinking tasks given to students. The importance of such structured yet flexible tools for assessing creative thinking in design-based learning has been emphasized in earlier work [10]. Previous studies have considered the design of rubrics for the assessment of design thinking tasks [11].

Assessment in design thinking must also account for interdisciplinary complexity. A previous study that reviewed assessment practices in interdisciplinary STEM education emphasized that effective tools must align with both the process and content of student learning [12]. They highlight the value of flexible rubrics and open-ended tasks that enable students to integrate knowledge across various domains. This is particularly important in poster artefacts that draw from technical, human-centered, and visual communication skills.

However, rubric-based assessments have limitations. Despite their structured format, they remain vulnerable to subjective interpretation, especially in visual or creative work. This concern raises the potential value of AI-assisted assessment, particularly generative AI systems capable of interpreting visual and textual information.

Recent research has begun empirically testing the reliability of GPT-based grading in design education. The recent study [16] demonstrated that a Custom GPT-4 model achieved acceptable inter- and intra-rater reliability (ICC > 0.7) when calibrated through rubric-based prompt engineering and exemplar referencing. Their iterative design-based process established a methodological foundation for subsequent studies, including the present work, which extends this approach to secondary-school design-thinking posters combining visual and textual elements.

Other studies have explored AI applications in feedback and formative assessment. For instance, TeacherGAIA is an AI-assisted chatbot that supports self-directed learning through carefully engineered prompts [13]. The tool enables students to reflect on their learning, identify knowledge gaps, and take initiative in monitoring their progress. Rather than offering static feedback, TeacherGAIA encourages iterative thinking and inquiry, making it well-suited to the reflective and process-oriented nature of design thinking. Another recent study also examined the integration of AI into assessment through a MOOC-based, AI-assisted flipped teaching model for EFL writing. The study found that AI-assisted approaches enhanced both teacher instructional practices and student writing outcomes. While not focused on visual content, this work supports the broader applicability of AI feedback in open-ended, process-oriented learning tasks. In a related study, researchers investigated the impact of an LLM-based chatbot, wisdomBot, on middle-school learners [14]. While students using the chatbot submitted more work and achieved higher scores, they showed shallower conceptual understanding, highlighting the importance of balancing efficiency with depth in AI-based tools. This finding is critical for design thinking, where quantity must not replace quality of insight.

Despite this progress, significant gaps remain. Previous research using machine-learning techniques for design-thinking assessment focused mainly on textual content [4]. More recent comparative analyses between GPT-4 and human graders in writing assessment found that GPT-4's feedback was generally consistent with human raters in holistic scoring but diverged when interpreting creativity and nuance [15]. Few studies have examined the use of generative AI to assess visual and process-based artefacts like posters. While tools like TeacherGAIA demonstrate the feasibility of AI-guided feedback, their application in high-stakes or summative assessment contexts, especially those requiring visual and process analysis, remains limited. Questions about validity, reliability, pedagogical alignment, and teacher acceptance remain largely unanswered.

This paper addresses this gap by investigating how AI-assigned and TA-assigned scores align with instructor judgments in the context of the evaluation of design thinking posters comprising visuals and text. We investigate how teachers perceive AI-assigned scores vis-à-vis TA-assigned scores.

## III. METHODOLOGY

### *A. Journey Map Assessment Task and Rubric*

For the purpose of this study, we adapted a design thinking homework assignment with the objective of creating a user journey map and its accompanying assessment rubric from the Design Thinking and Innovation (DTI) module from Singapore University of Technology and Design (SUTD). In design thinking, a journey map is a visual representation of a user's experience as they interact with a product, service, or system. It captures the end-to-end process from the user's perspective, highlighting their actions, emotions, pain points, and opportunities for improvement. Journey maps are used to build empathy and uncover insights that inform human-centered solutions. Visuals in the journey map enhance clarity and storytelling. The journey map from the DTI module was measured on three main criteria (Fig. 1):

1. **Empathy and user understanding**: demonstrated by capturing the user's emotions and experiences.
2. **Identification of pain points and opportunities**: Demonstrated by outlining moments of difficulty that the user experiences during their interaction with a product, service, or system and potential moments for improvement or innovation based on those pain points or unmet user needs.
3. **Visual communication**: use of layout, imagery, and design elements to clearly and effectively convey information.

The rubric for the journey map assessment task was developed and validated by faculty members at SUTD.

### B. *Selection of Journey Map Samples*

The research team selected journey map student samples submitted in the DTI module at SUTD (Fig. 2). The selection process aimed to ensure a representative range of performance across the three rubric dimensions: empathy and user understanding, identification of pain points and opportunities, and visual communication. Samples were chosen to reflect varying levels of visual layout, such as those that are (a) structured, (b) semi-structured, and (c) free flowing, to facilitate meaningful comparisons across AI, TA, and teacher assessments. All samples were anonymized. For task 1, ten samples were selected to test the alignment between AI and teacher scores. For Task 2, a different set of ten samples was curated to avoid bias from prior exposure and to support randomized presentation in the teacher confidence task.

### C. *Development of Journey Map Grader*

The journey map grader used in this study was developed on GPT-4o (OpenAI, May 2024 release) to evaluate student journey maps based on a structured rubric (Fig. 1). The development process of the journey map grader followed the approach detailed in [16], which demonstrated how a Custom GPT-4 model can be iteratively calibrated for reliable design-assignment grading through prompt engineering and exemplar referencing. Rather than performing full model retraining, the grader in this study was prompt-engineered and calibrated using ten student samples from the DTI module that had been previously scored by teaching assistants. These human-assigned scores were used to calibrate the AI's scoring logic with the rubric descriptors, ensuring transparent and rule-based evaluation. To verify consistency, the same samples were re-uploaded across multiple runs, and the generated scores were compared to confirm output stability.

### D. *Scoring by Teaching Assistants, Journey Map Grader and Development of Hybrid Scores*

The selected samples were scored by teaching assistants at SUTD. The teaching assistants (TAs) scored the student journey map samples using a structured rubric that assessed three key dimensions: empathy and user understanding, identification of pain points and opportunities, and visual communication. The instructor's original scores from the DTI module served as the benchmark reference for comparison with AI- and TA-assigned scores. To facilitate consistent and collaborative evaluation, a sample board was set up on Miro, where TAs and faculty members could view and discuss the samples in real time. This setup allowed for clarification of rubric criteria and alignment of scoring interpretations across assessors. TAs independently marked the samples, and discussions were encouraged to resolve ambiguities and ensure fairness. Once the initial scoring was completed, the final set of marked samples was reviewed by an instructor to validate the scores and ensure consistency with pedagogical expectations. This multi-layered process helped maintain reliability and rigor in the assessment of visual and process-based design thinking artefacts.

The research team utilized the journey map grader to generate the scores for each sample. Hybrid scores were computed as the average of AI-assigned and TA-assigned scores for each rubric dimension.

| | 1<br>MINIMAL | 2<br>DEVELOPING | 3<br>SATISFACTORY | 4<br>STRONG | 5<br>EXCEPTIONAL |
|---|---|---|---|---|---|
| **Empathy and User Understanding**<br><br>how well the journey map demonstrates a deep understanding of the user's emotions, needs, and experiences. | Minimal or no empathy; user needs and emotions are absent or unclear. | Limited empathy; user motivations and emotions are vaguely addressed or inconsistently represented. | Basic empathy; user needs and emotions are identified but lack depth or comprehensive understanding. | Clear empathy; user emotions, motivations, and needs are well-detailed but may lack minor nuances or diverse perspectives. | Profound empathy; user emotions, motivations, and needs are vividly captured, reflecting a thorough understanding of diverse user perspectives. |
| **Identification of Pain Points and Opportunities** | Pain points and opportunities are missing or poorly articulated. | Limited identification of pain points or opportunities, with weak connections to user needs. | Basic pain points and opportunities are identified but are superficial or underdeveloped. | Pain points and opportunities are identified clearly but may lack depth or minor alignment issues with user needs. | Pain points are precisely identified with insightful opportunities for improvement that are well-aligned with user needs. |
| **Visual communication** | Visuals are absent or ineffective, hindering understanding of the journey map. | Visuals are poorly executed, cluttered, or do not effectively support the narrative. | Visuals are functional but basic, with limited engagement or impact. | Visuals are clear, appealing, and support the narrative well, with minor room for improvement. | Visuals are professional, highly engaging, and effectively enhance the narrative and usability of the journey map. |

Fig. 1. Rubric used to train the journey map grader

*E. Study Setup*

The study was conducted as part of the Design Centric Interdisciplinary Creative Problem-Solving (ICPS-D) teacher workshop held in June 2025, involving secondary school teachers from the MOE, Singapore. Teachers had been introduced to the journey map method and had utilized it for their own design process. The workshop format allowed for rich interaction, immediate feedback, and reflection on the use of AI in formative assessment, making it an ideal environment to pilot the study's methodology and gather insights from experienced educators.

*F. Participant Background*

A total of 33 in-service teachers from MOE, Singapore secondary schools (n = 29) and pre-university institutions (n = 4) participated in this study. Most were mid-career educators aged between 40 and 49 years (n = 19, 57.6%), with the majority having more than 10 years of teaching experience (n = 28, 84.9%). Participants represented a diverse range of subject areas, with many teaching across multiple disciplines. The most common subjects among participants were Chemistry and Physics (each n = 7, 13.5%), followed by Science (n = 6, 11.5%) and Mathematics (n = 5, 9.6%). Other subjects included History, Biology, Social Studies, Design and Technology, Project Work, and General Paper.

*G. Experimental Activities*

Teachers were given two tasks. In the first task, they were provided with ten anonymized journey map samples and asked to evaluate them manually using the given rubric, i.e., by interpreting the rubric independently and entering the scores in a Google Form. The presentation order of samples was randomized for each participant. Participants were free to discuss the interpretation of the rubric to establish benchmarks, but they performed the scoring independently. Teachers were given 45 minutes to complete this task.

The objective of the second task was to evaluate teachers' agreement with three different scoring groups: AI-based, Teaching assistants (TAs), and a hybrid score combining both. A different set of ten samples was chosen to prevent any bias from the first task. Each sample was presented with three score options (AI, TA, Hybrid), in randomized order in an online form. The three score options were unlabeled, so teachers did not know which score belonged to which group.

Teachers were asked to indicate which of the three score groups they agreed with the most. The task was completed within a 45-minute session and followed by a survey with open-ended questions. To enrich the findings, an open focus group-style discussion was held at the end of the workshop, where teachers shared their views on the role of AI for assessment of such tasks and its potential integration into classroom practice.

*H. Analyses*

The study used both quantitative and qualitative methods. Quantitatively, it measured how much AI, TA, and instructor scores agreed, using percentage agreement and Cohen's Kappa, a standard measure of inter-rater reliability that adjusts for chance agreement. Given the exploratory scale, we reported descriptive reliability measures (percentage agreement, Cohen's Kappa). Qualitatively, it analyzed open-ended responses and focus group discussions to identify themes about teachers' attitudes toward AI-assisted assessment. This combined approach captured both the statistical alignment of scores and the factors shaping teachers' perception of AI grading using the journey map grader.

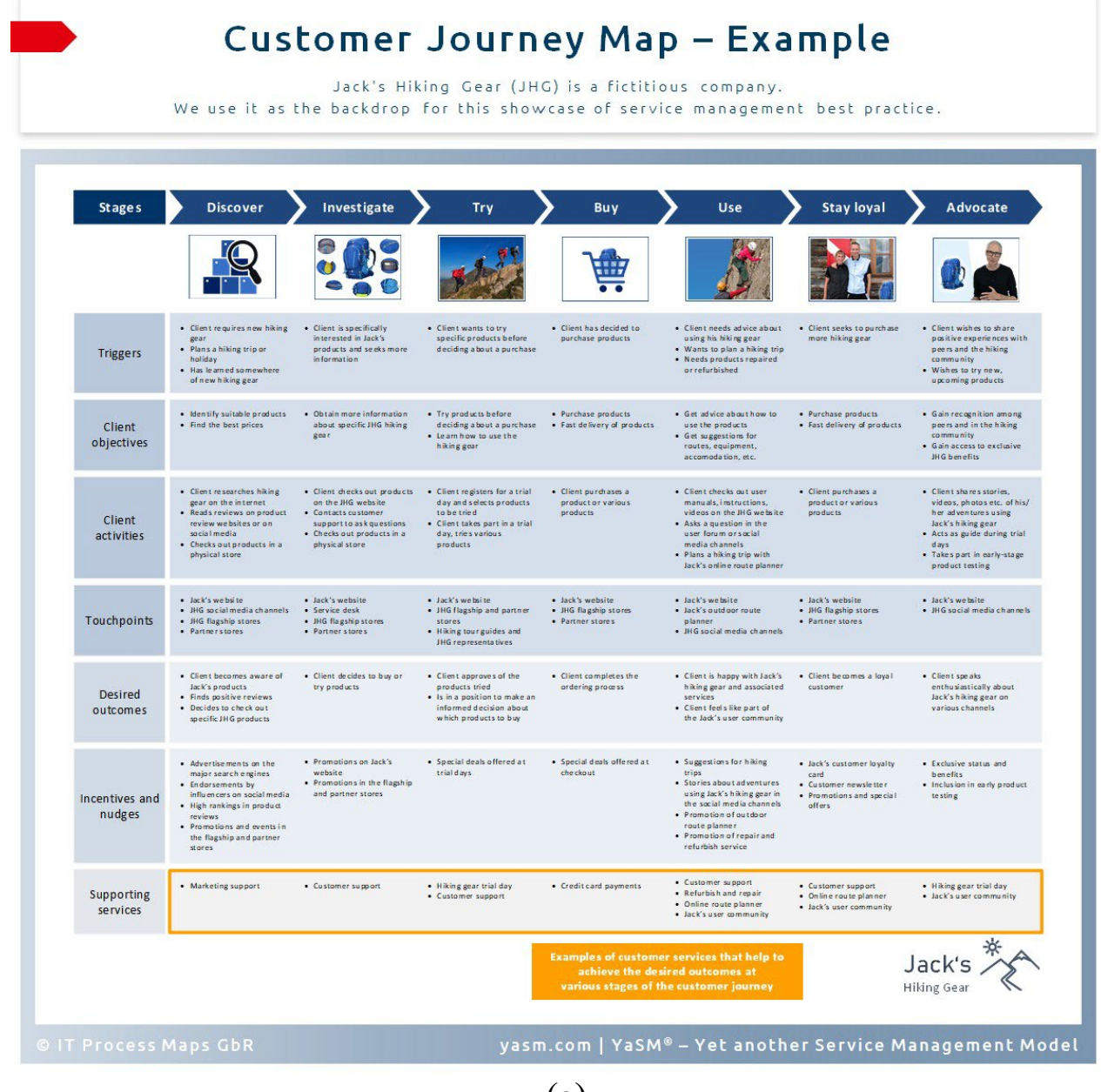

(a)

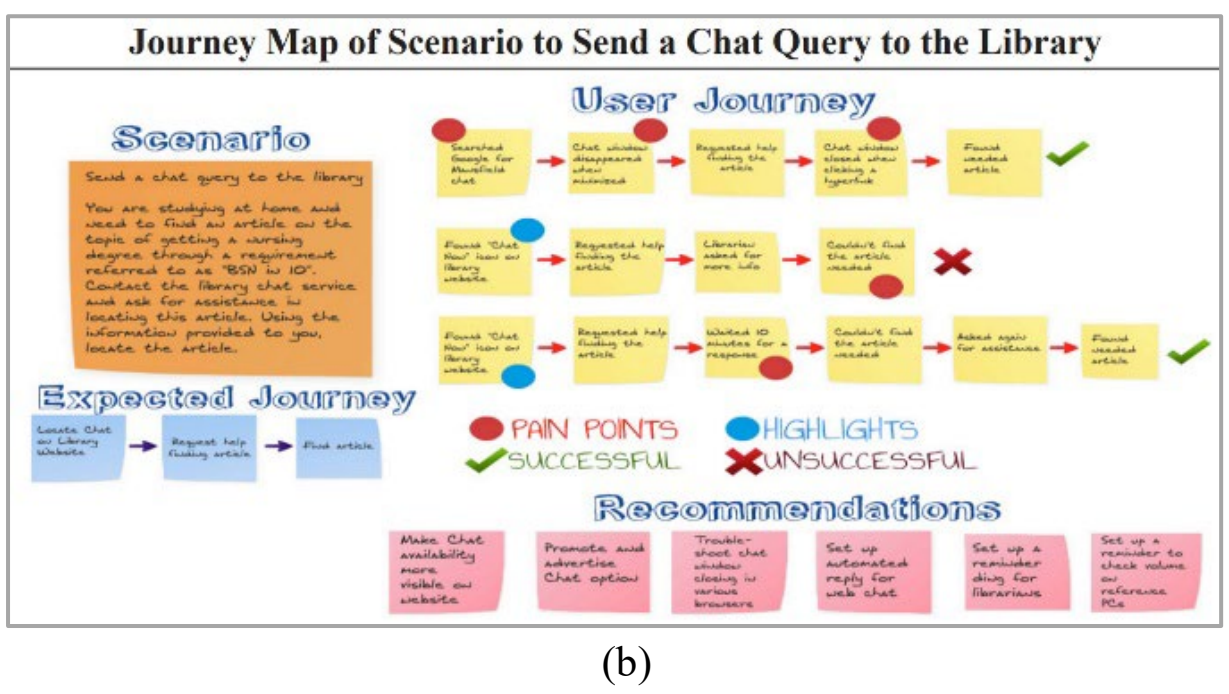

(b)

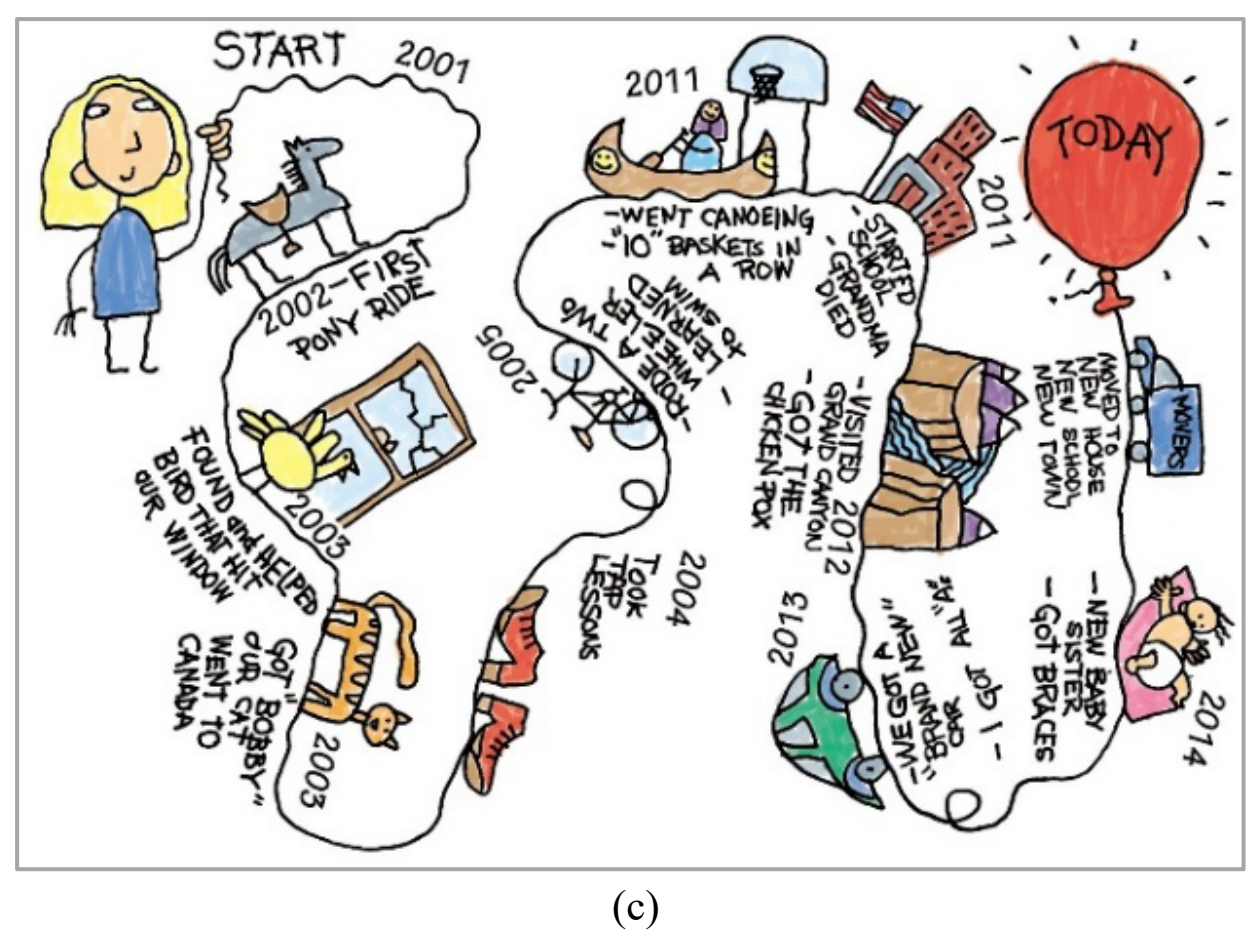

(c)

Fig. 2. Examples of journey maps similar to students samples used in study (a) Structured journey map (b) Semi-structured journey map (c) Free flowing journey map * *All above images by Unknown Authors are licensed under CC BY-ND*

## IV. Results

### A. Comparison of AI-assigned Scores, TA-assigned Scores and Teacher-assigned Scores

In the first task, teachers evaluated ten anonymized student journey maps using a structured rubric manually. Average scores assigned by teachers were compared with those generated by the journey map grader and TAs to assess agreement between the grading groups. (Fig. 3). For this exploratory study, we used percentage agreement and Cohen's Kappa to compare ratings provided by different groups.

For the dimension Empathy and User understanding, Instructor scores ranged from 2.0 to 4.5, with noticeable variation across samples. AI scores were consistently around 4.0, showing minimal variance and little alignment with instructor fluctuations. TA scores fluctuated more significantly, aligning more closely with instructor scores in certain samples (e.g., 1-b, 1-h), but diverging in others (e.g., 1-g, 1-i). Instructor vs AI showed only 20% agreement, with a negative Kappa (–0.026), indicating a clear divergence in scoring patterns. Instructor vs TA had 30% agreement, with a Kappa of 0.091, reflecting slight agreement (Table I).

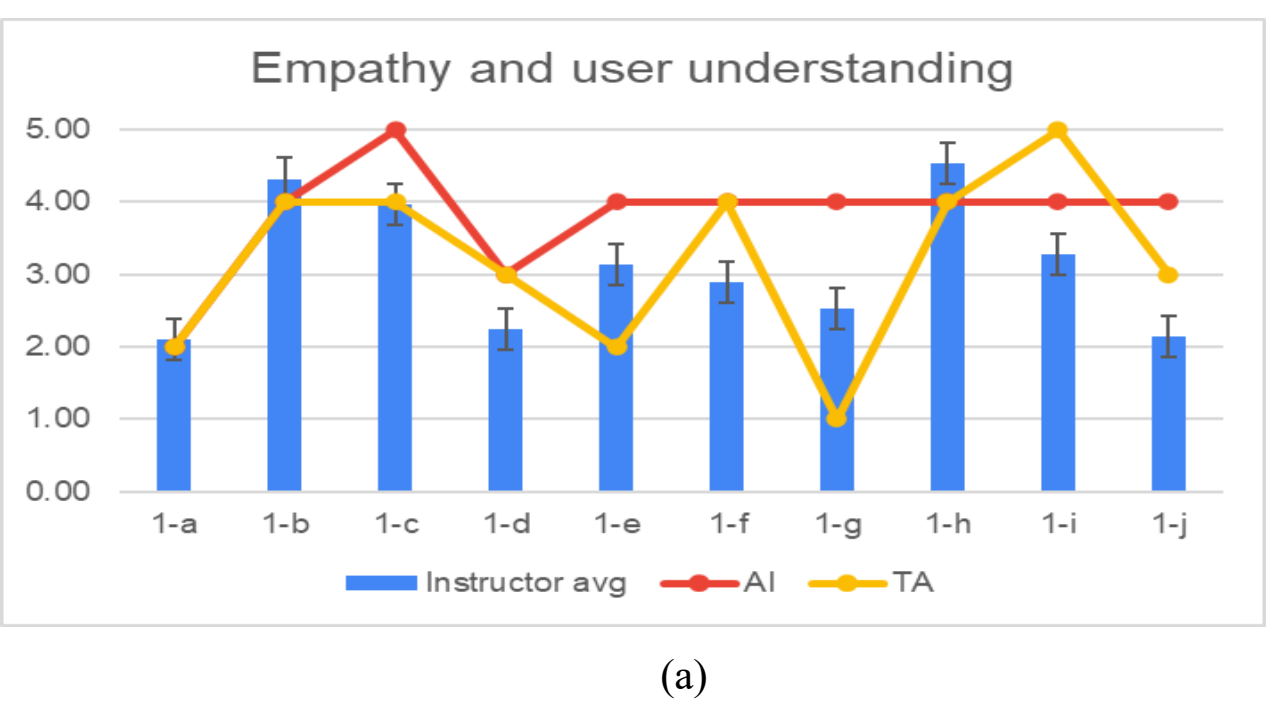

(a)

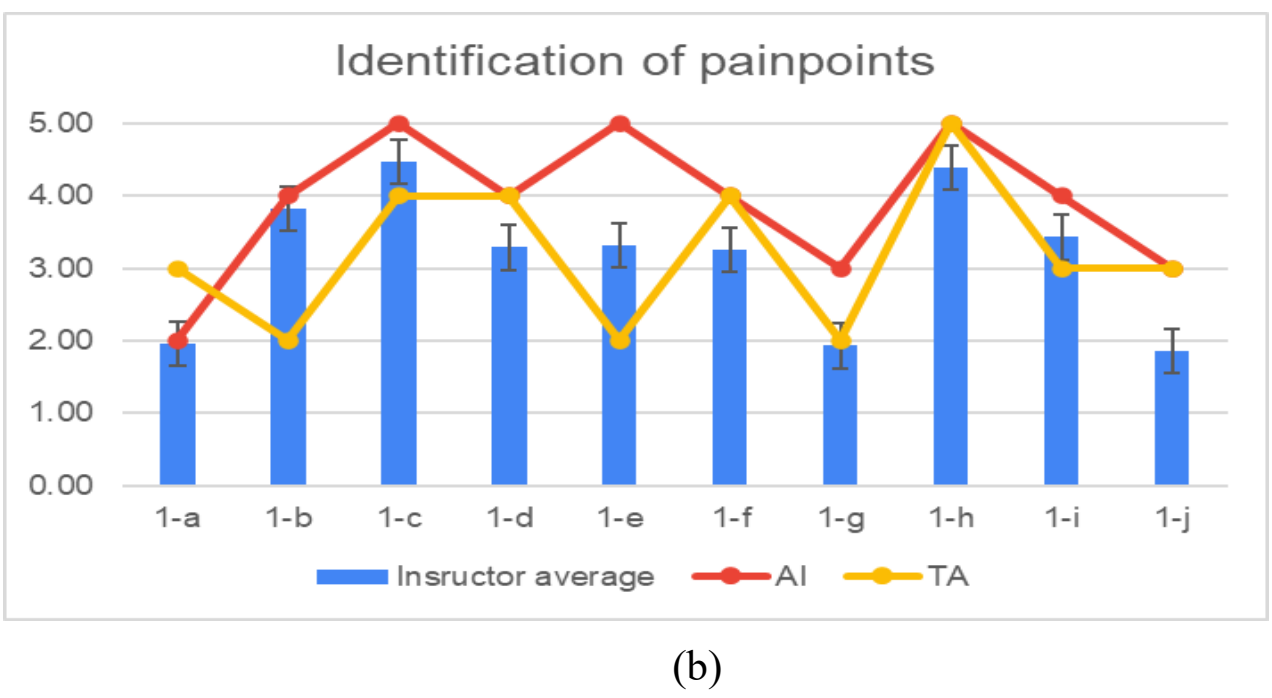

(b)

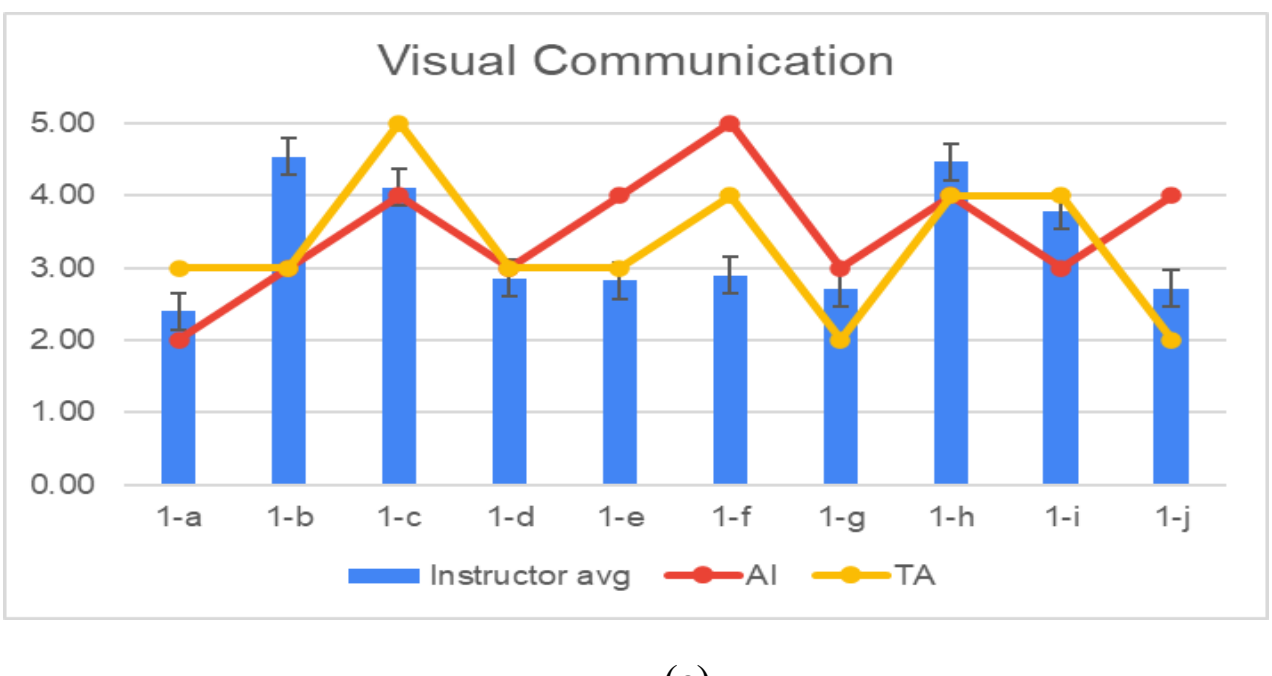

(c)

Fig. 3. Teachers' average scoring of samples, with AI and TA-assigned scores (a) Empathy and user understanding, (b) Identification of pain points and opportunities, (c) Visual Communication

For the dimension Identification of pain points and opportunities, instructor averages show scoring between 2.0 and 4.5. Comparisons involving the instructor ratings showed low agreement: Instructor vs AI had only 20% raw agreement and a negative Kappa (–0.039). Instructor vs TA had 30% raw agreement, but Kappa = 0.000, indicating no agreement beyond chance, with p = 1.000 (Table II).

For the Visual Communication rubric dimension, Instructor scores show variation across samples, particularly higher scores in 1-b and 1-h. AI scores are relatively stable (mostly 3.0–4.5), again lacking variation. TA scores exhibit broader fluctuations, particularly peaking at 1-c and 1-h, aligning more with instructor highs. The highest raw agreement was observed between the instructor and AI (50%), followed by the instructor and TA (40%). Cohen's Kappa values showed a similar pattern, with fair agreement between instructor and AI (κ = 0.242, p = .243), slight agreement between instructor and TA (κ = 0.118, p = .551) (Table III).

### B. Teachers' Perception of AI-assigned Scores, TA-assigned and Hybrid Scores

In the second task, teachers reviewed a different set of ten student samples, each accompanied by three anonymized scores: AI-assigned, TA-assigned, and Hybrid scores, and were asked to select the score they agreed with the most. Teachers showed a preference for TA-assigned scores in six of the ten samples (Fig. 4). In four samples, more teachers preferred AI-assigned scores than TA-assigned or hybrid scores. Hybrid scores were not a top choice for any of the samples.

### C. Analysis of Qualitative Comments by Teachers

Open-ended survey responses and focus group discussions revealed that while teachers appreciated the speed and consistency of AI grading, they expressed concerns about its ability to capture nuanced aspects of student work, particularly empathy and contextual insight. Teachers expressed a range of perspectives on the use of AI-assisted scoring compared to manual grading of journey maps.

#### 1) Advantages of AI

A recurring theme was the efficiency and speed of AI-based scoring, which many found helpful, especially for formative feedback. Several teachers appreciated that the

TABLE I Comparison of Empathy scores

| Comparison | % | Kappa | Interpretation | p |
|---|---|---|---|---|
| Instructor vs AI | 20% | –0.026 | Less than chance | 0.848 |
| Instructor vs TA | 30% | 0.091 | Slight agreement | 0.574 |

a. % = % Agreement, p = p-value

TABLE II Comparison of Identification of pain points & opportunities scores

| Comparison | % | Kappa | Interpretation | p |
|---|---|---|---|---|
| Instructor vs AI | 20% | –0.039 | Less than chance | 0.799 |
| Instructor vs TA | 30% | 0 | No agreement | 1 |

a. % = % Agreement, p = p-value

TABLE III Comparison of Scores for visual communication

| Comparison | % | Kappa | Interpretation | p |
|---|---|---|---|---|
| Instructor vs AI | 50% | 0.242 | Fair agreement | 0.243 |
| Instructor vs TA | 40% | 0.118 | Slight agreement | 0.551 |

a. % = % Agreement, p = p-value

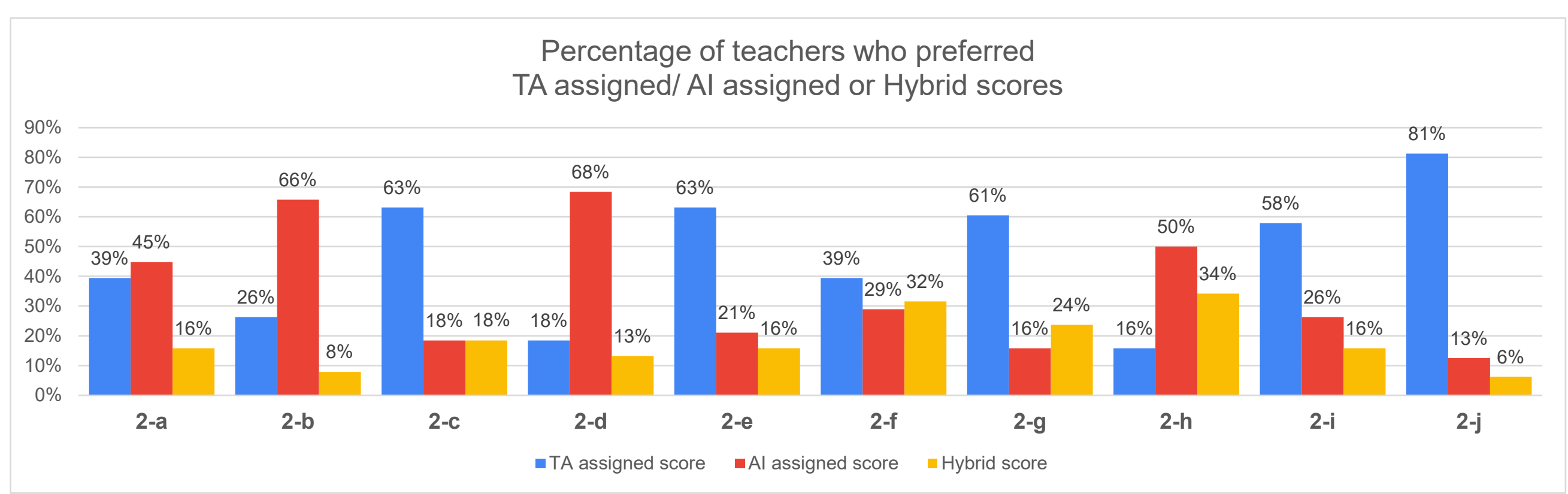

Fig. 4. Percentage of teachers who preferred TA-assigned, AI-assigned or hybrid scores

journey map grader could provide rationale for its scores and saw potential for it to support student self-directed learning by offering immediate, iterative feedback without requiring teacher intervention.

*2) Concerns about AI*

However, concerns about accuracy and trust were prominent. Teachers noted that while some AI-generated scores aligned closely with their own, others diverged significantly—particularly in cases involving nuanced visual interpretation. Some samples that teachers rated low were scored more generously by the AI, raising doubts about its reliability in high-stakes contexts.

*3) Concerns about Manual*

Most teachers described manual scoring as slow, tedious, labor-intensive, and time-consuming, as it required extensive discussion among team members, both before the grading to establish benchmarks as well as during the grading process to discuss variants. Teachers acknowledged the challenges of interpreting visual content and applying rubrics consistently. The issue of subjectivity towards visual content cropped up as a theme, and questions were raised about different assessors interpreting the rubrics differently, due to their experience, preferences of aesthetic appreciation, or 'personal emotions' while grading.

*4) Advantages of Manual Methods*

Nevertheless, several teachers appreciated the deeper engagement and discussion that manual scoring enabled, which helped clarify rubric expectations and fostered a better understanding of student work. Several teachers saw manual scoring as more accurate, fairer, and authentic, as 'there was more connection and understanding of the content being graded.'

*5) Surprising*

Interestingly, one teacher pointed out that manual rubric-based scoring by human assessors could also have inconsistencies, whereas AI scoring could be consistent for a large student cohort (Participant 17). Another teacher agreed that AI-based scoring had less bias (Participant 2). Finally, while teachers appreciated the speed of AI-based scoring, there was concern about the setup time for an accurate and consistent grader. As articulated by participant 18,

*"The manual scoring one is more labor-intensive and has a lot of subjectivity depending on the assessor. Comparing different members of our group, we found that we had instances of differences in the standards applied for scoring. The AI-based scoring is much faster, with a rationale given for the scores assigned. However, we wonder if time has to be spent to train the AI first for consistent and accurate scoring effectively."*

Finally, teachers suggested that AI could be more effective if it were trained to mimic individual teacher marking styles. However, others were skeptical about whether AI could truly capture the nuances of creativity and visual reasoning. Overall, teachers saw promise in AI as a complementary tool, useful for efficiency and formative feedback, but not yet a replacement for human judgment in summative assessment. As commented by a participant,

*"I've always been thinking about how I can use AI not just as an answer provider but as a buddy for teaching or learning. So, for the case of the students, I like the part where you share the graders with your students so that they don't always have to see us to give them feedback. So, they get feedback chatbot, maybe improve on their own based on how much they want to work on. So if they work on it, then they will get a better grade. It is very self-directed in nature, which I think lets the student decide what they want to do."*

## V. DISCUSSION AND CONCLUSION

We presented an exploratory study investigating the reliability and perceived accuracy of AI-assisted and TA-assisted assessment in evaluating student works, such as posters that comprise both textual and visual elements. Findings from experimental tasks conducted with teachers from MOE, Singapore secondary schools and pre-university institutions offer valuable insights into the potential and limitations of AI-based grading in public education systems.

In the first task, AI scoring appeared less responsive to variations in student performance, possibly indicating a generic evaluation pattern. As seen in the empathy dimension, AI scoring remained clustered around mid-to-high values despite instructor variation, suggesting a tendency toward generic evaluation patterns and surface-level sensitivity to emotional or structural cues rather than authentic user reasoning. In the identification of pain points, in particular, AI tended to over-score student work, potentially lacking nuanced recognition of deficiencies. Discrepancies in scoring highlight the importance of contextual interpretation, especially in assessing abstract qualities such as empathy and visual communication, which AI may struggle to fully capture without deeper contextual input. While AI could replicate human judgment to a reasonable extent, it required further calibration with rubric-based training data. The low Cohen's Kappa values across empathy and pain-point dimensions underscore the challenge of automating assessments where human interpretation is inherently contextual. Such divergence highlights potential risks in deploying AI graders for high-stakes evaluation, particularly bias toward surface-level textual or visual cues rather than interpretive depth.

The second task provided deeper insight into teacher perceptions. Teachers expressed a preference for TA-assigned scores in six out of ten samples, while AI-assigned scores were preferred in four samples. Teachers consistently place more trust in TA scoring, with slightly less enthusiasm for AI scoring. This suggests that while automation can support evaluation, human judgment remains essential, particularly in subjective or design-oriented assessments.

These findings are consistent with emerging research comparing AI and human grading in creative or open-ended learning contexts. For example, [16] reported that a Custom GPT-4 grader achieved high inter- and intra-rater reliability (ICC > 0.7) in a university design-studio context when calibrated with rubrics and exemplar prompts. At the same time, [15] found that GPT-4's feedback was generally consistent with human raters on holistic quality but diverged on creative and contextual interpretation. The present study extends these insights to secondary-school design thinking assessment, where lower inter-rater agreement (Kappa < 0.25) suggests that task complexity and multimodal artefacts may limit the direct transferability of GPT reliability across educational levels. Together, these parallels indicate that while generative AI can enhance scalability and consistency, contextual interpretation and pedagogical alignment remain crucial for credible assessment in creative disciplines.

Our results reflect how each scoring method aligns with teachers' expectations in terms of evaluation reliability and perceived accuracy. Assessment by teaching assistants, who are often junior teachers or graduate students, brings certain limitations. While they may possess foundational knowledge and a degree of pedagogical training, their evaluation of student work can be inconsistent, coloured by limited experience and varying interpretations of assessment rubrics. AI graders promise consistency and efficiency, rapidly processing large volumes of student submissions with uniform application of criteria. However, AI systems, although robust in terms of scalability and speed, often lack the nuanced understanding required to interpret creative or abstract elements in student work. This can lead to a mechanical application of scoring guidelines, potentially overlooking the subtleties of individual student expression.

Both approaches, therefore, present distinct challenges: human assessors risk variability and subjectivity, while AI systems, although impartial, may fall short in capturing the richness and complexity inherent in creative assignments. As such, neither method alone fully addresses the multifaceted demands of fair and comprehensive assessment in educational contexts. As the hybrid scores utilized in this study were based on averages and were thus preliminary, they were viewed less favorably by teachers, as they lacked contextual grounding. Further research is necessary to develop hybrid assessment approaches that integrate computational efficiency and consistency with human insights.

Beyond methodological considerations, AI-assisted assessment also raises important ethical challenges. Issues such as algorithmic bias, data privacy, and transparency in automated scoring must be carefully managed to ensure fairness and accountability. Future implementations should include human oversight and institutional safeguards to uphold ethical standards when integrating AI into assessment processes.

### A. *Recommendations for Development and Adoption of AI-Based Assessment in Creative Fields*

Based on the insights presented in the paper, we present recommendations for the adoption of AI-based assessment for design assessment in public educational settings:

*1) Development of Hybrid Assessment Models:*
Given the limitations of both AI and human scoring, especially in creative domains, hybrid models should be researched and developed further. AI can perform first-pass evaluations by efficiently processing large volumes and offering immediate feedback, while teachers or TAs moderate and refine scores. This balances speed and scalability with the contextual judgment essential for fair assessment.

*2) Aligning AI training with Instructor Scoring Patterns:*
The success of AI graders relies heavily on the reliability of the bot for evaluating student works. More research is required for the development of reliable bots that facilitate updating the training data, refining the scoring algorithms, and incorporating new assessment criteria as needed. Teachers should also be able to fine-tune models based on individual teacher styles or disciplinary expectations. Calibration using annotated training data from experienced instructors can significantly enhance both alignment and teacher trust in AI scoring.

*3) Teacher Readiness and Training:*
Our findings indicate that teachers are cautiously optimistic about integrating AI into their assessment workflows. To enhance teacher readiness, professional development programs should be implemented to train educators on how to effectively use AI tools. This includes understanding the AI's scoring logic, interpreting AI-generated feedback, and integrating it with their own assessments.

*4) AI Graders for Formative Feedback and Self-Evaluation of Design Thinking Assessment:*
Until more reliable graders are developed, AI graders should be positioned as tools for formative assessment rather than final grading. This approach allows students to use AI feedback to improve their work iteratively. By providing immediate, detailed feedback, AI can help students identify areas for improvement and engage in self-directed learning. This is aligned to the recommendation by [16] that highlighted the potential of generative AI for providing feedback in open-ended, creative tasks.

While the study offers promising insights into the use of AI-assisted assessment in design education, several limitations must be acknowledged. First, the sample size of both the journey maps and participating teachers was relatively small, constraining the generalizability of the findings. The number of artefacts evaluated per task was limited to ten, which may not fully capture the diversity of creative student work in large cohorts. Second, the journey map grader was calibrated on a specific rubric and dataset, which may have influenced its scoring performance. Third, teacher perceptions were collected in a single workshop setting; longitudinal studies are needed to examine how perceptions of AI grading evolve with sustained use. Future work should refine AI scoring mechanisms, improve rubric alignment, and develop more advanced hybrid models. Researchers should also incorporate a larger dataset of participants and artefacts to enhance robustness. Expanding the range of artefact types and examining long-term classroom integration would further deepen understanding of AI graders' practical and pedagogical impact.

This study investigated the potential of AI-assisted assessment for evaluating student works, such as posters that comprise textual and visual elements, in design thinking education. By comparing AI-, TA-, and teacher-assigned scores and examining teachers' perceptions of different scoring groups, the research highlights the value of understanding both the technical alignment and the human acceptance in automated scoring systems. As design education continues to grow in scale and complexity, thoughtfully integrated AI tools can enhance assessment practices and reduce grading burdens. By surfacing teacher attitudes and evaluating alignment across scoring systems, this work contributes to the evolving discourse on responsible AI adoption in creative disciplines, highlighting the promise of hybrid models that combine automation with human insight to support scalable and pedagogically sound assessment.

## ACKNOWLEDGMENT

This research was supported by the Ministry of Education (MOE), Singapore under the Education Research Funding Programme (ERFP 21/23 RA grant), administered by by the Singapore University of Technology and Design (SUTD).

## AI DISCLOSURE

This manuscript includes content drafted and refined with the assistance of OpenAI's ChatGPT. All conceptual development, analyses, and interpretations were conducted and validated by the authors.